\documentclass[12pt]{article}
\usepackage{amsmath,amssymb,amsfonts,color,graphicx,cite,color}
\usepackage{latexsym}
\usepackage[normalem]{ulem}
\usepackage{epsfig,psfrag,rotating,soul}
\usepackage{rotfloat}
\usepackage{tablefootnote}
\usepackage{orcidlink}
\input paperdef
\evensidemargin \oddsidemargin
\allowdisplaybreaks

\begin{document}
\thispagestyle{empty}

\def\thefootnote{\fnsymbol{footnote}}

\begin{flushright}
\mbox{}
IFT--UAM/CSIC--24-149 
\end{flushright}

\vspace{0.5cm}

\begin{center}

\begin{large}
\textbf{Lepton Flavor Violation in
  Nonholomorphic Soft SUSY-Breaking\\[2ex]
  Scenarios: Experimental Limits and Excesses}
\end{large}

\vspace{1cm}

{\sc 
M.~Rehman\orcidlink{0000-0002-1069-0637},$^{1}$%
\footnote{email: m.rehman@comsats.edu.pk}%
~and S.~Heinemeyer\orcidlink{0000-0002-6893-4155},$^{2}$%
\footnote{email: Sven.Heinemeyer@cern.ch}%
}

\vspace*{.7cm}
{\sl
${}^1$Department of Physics, Comsats University Islamabad, 44000
  Islamabad, Pakistan \\[.1em] 
${}^2$Instituto de F\'isica Te\'orica, (UAM/CSIC), Universidad
  Aut\'onoma de Madrid, Cantoblanco, 28049 Madrid, Spain
}
\end{center}

\vspace*{0.1cm}

\begin{abstract}
\noindent

We investigate the prospects for the observation of lepton flavor
violation (LFV) within the nonholomorphic supersymmetric standard model
(NHSSM). We examine charged lepton flavor-violating (cLFV) decays such
as such $\mu \rightarrow e \gamma$, $\tau \rightarrow e \gamma$ and
$\tau \rightarrow \mu \gamma$ to impose indirect constraints on both LFV
holomorphic and nonholomorphic (NH) soft Supersymmetry-breaking (SSB) terms.
These constraints are subsequently utilized to calculate decay
rates for LFV Higgs decays (LFVHD). Within the allowed parameter space,
NH contributions to LFVHD can be notably larger compared to the
holomorphic counterparts.
Interestingly, recently ATLAS reported an excess larger than
$2\,\si$ in their searches for $h \to e \tau$ and $h \to \mu\tau$.
Their best-fit point is not excluded by the corresponding CMS limits.
We demonstrate that for some parts of parameter space, the
predicted values for $\br(h\to e \tau)$ and $\br(h\to \mu \tau)$ in the
NH scenarios reach up to the 
present experimental limit for these decay processes and can
potentially explain the excesses observed by ATLAS, while being in agreement
with other experimental constraints. If these decays
are eventually observed experimentally, they could
potentially serve as a distinctive signature of the NH scenarios
and determine some of the NH parameters.
Conversely, the limits on the LFVHDs can
restrict the allowed parameter space for the NH SSB terms in the NHSSM. 

\end{abstract}

\def\thefootnote{\arabic{footnote}}
\setcounter{page}{0}
\setcounter{footnote}{0}

\newpage



\section{Introduction}
\label{sec:intro}

Neutrino oscillations~\cite{Super-Kamiokande:1998kpq, SNO:2001kpb,
  SNO:2002tuh} serve as a compelling indication of physics beyond the
Standard Model (SM). An intriguing consequence of neutrino oscillations
is the existence of lepton flavor violation (LFV), which is not
explicitly allowed in the SM as the lepton flavor conservation emerges
as an incidental symmetry\cite{Isidori:2021gqe}. As LFV manifests itself
in the neutrino sector, there is a reasonable expectation that it may
also occur in the charged lepton sector. LFV is strictly prohibited in
the SM and its simple extensions predict negligible rates for charged
lepton flavor violation (cLFV) decays such as $\mu \rightarrow e
\gamma$, $\tau \rightarrow e \gamma$ and $\tau \rightarrow \mu \gamma$
\cite{Marciano:1977wx, Cheng:1980tp}, rendering it beyond the reach of
current experiments. Conversely, new physics models\cite{Hisano:1995nq,
  Ellis:2002fe,Sher:2002ew} offer the possibility of predicting cLFV at
levels that can be detected by existing and future low-energy
experiments. This, in turn, offers a way to constrain and assess
scenarios involving new physics.      

The Minimal Supersymmetric Standard Model
(MSSM)~\cite{Fayet:1974pd,Fayet:1976et,Fayet:1977yc, Nilles:1983ge,
  Haber:1984rc,Barbieri:1987xf} stands out as one of the most attractive
and extensively studied new physics models. Beyond addressing lingering
questions posed by the SM, the MSSM introduces an additional source of
LFV through soft supersymmetry-breaking (SSB)
terms~\cite{Hall:1985dx,Borzumati:1986qx}. This additional degree of
freedom holds the potential to yield predictions for cLFV decays at
levels comparable to the existing experimental constraints on cLFV
decays\cite{Masina:2002mv, Paradisi:2005fk,Crivellin:2010er,
Crivellin:2011jt,Arana-Catania:2013nha}. However, the conventional
structure of the MSSM is increasingly constrained by the results of the
Large Hadron Collider (LHC)\cite{Sekmen:2022vzu},
prompting researchers to explore alternatives or
modifications of the MSSM. One approach involves the incorporation of
nonholomorphic (NH) SSB terms~\cite{Girardello:1981wz, Bagger:1993ji},
leading to a modified model known as the nonholomorphic supersymmetric
standard model (NHSSM) \cite{Cakir:2005hd}.
The additional NH terms can lead to notable phenomenological
outcomes that help to align the SUSY predictions with current
experimental data. As an example, the NHSSM has the potential to
significantly reduce electroweak
fine-tuning~\cite{Ross:2016pml,Ross:2017kjc}.
Radiative corrections to the
Higgs mass are influenced by both holomorphic and NH trilinear SSB
parameters of the third generation squarks, leaving additional possibilities
to more easily satisfy the Higgs-boson mass
constraint~\cite{Rehman:2022ydc}. Additionally, the model provides
parameter space compatible with the $\br(B \to X_s + \gamma)$ constraint
at high $\tb$, a part of parameter space that within the MSSM without NH
soft terms is more readily constrained~\cite{Chattopadhyay:2016ivr}.

The NHSSM has already been studied in detail~\cite{Un:2014afa, Jack:1999ud,
  Jack:1999fa, Jack:2004dv, Sabanci:2008qp,
  Israr:2024ubp}, with recent attention directed towards its LFV
aspects, as discussed in \citere{Chattopadhyay:2019ycs}. The latter
study proposed that NH terms could potentially predict branching ratios
for LFV Higgs decays (LFVHD) that are two orders of magnitude larger
compared to the predictions based solely on holomorphic SSB
terms. However, the analysis in \citere{Chattopadhyay:2019ycs} employed
a multi-parameter scan approach, lacking a precise quantification of the
specific contributions of NH terms to LFV decays. In contrast, our
approach in this paper involves systematically altering one (or two)
parameter(s) at
a time, providing a clear indication of the NH contributions. As
outlined in the following sections, this method yields predictions for
LFVHD that differ from those already available in the
literature. Alongside offering these refined predictions for LFVHD, we
also present updated constraints on both holomorphic and
NH SSB terms, originating from cLFV decays and LFVHD.

Interestingly, recently ATLAS reported an excess~\cite{ATLAS:2023mvd}
in their searches for $h \to e \tau$ and $h \to \mu\tau$. Together these
two channels show an excess larger than $2\,\si$. Their best-fit values
of $\br(h \to e \tau) \approx \br(h \to \mu\tau) \approx 0.1\%$ is not
excluded by the corresponding CMS limits~\cite{CMS:2021rsq},
$\br(h \to e\tau) < 0.22\%$ and $\br(h \to \mu\tau) < 0.15\%$ at the
95\% C.L.~limit. We demonstrate that the NHSSM can accommodate both
excesses without being in conflict with other experimental limits.

For our analysis, we developed a
{\tt SPheno}~\cite{Porod:2003um} source code using the Mathematica
package {\tt
  SARAH}~\cite{Staub:2009bi,Staub:2010jh,Staub:2012pb,Staub:2013tta,Staub:2015kfa}. The
{\tt SARAH Scan and Plot} ({\tt SSP}) \cite{Staub:2011dp} package served as an
interface to {\tt SPheno} for conducting numerical calculations. 

The paper is organized as follows: first, we review the essential
features of the NHSSM in \refse{sec:model_NHSSM}. The computational
framework is explained in \refse{sec:CalcSetup} whereas the numerical
results are presented in \refse{sec:NResults}, including our
analysis of the ATLAS excesses in their searches for $h \to e\tau$ and
$h \to \mu\tau$. Finally, our conclusions
can be found in \refse{sec:conclusions}.


\section{Model set-up}
\label{sec:model_NHSSM}

The necessity for holomorphicity in the superpotential within the MSSM results in the use of holomorphic operators to parameterize the SSB sector. However, it is possible to extend the MSSM by introducing terms that violate R-Parity and/or incorporate NH terms in the SSB sector~\cite{Girardello:1981wz, Bagger:1993ji, Chakrabortty:2011zz}. In its most basic formulation, the NH SSB sector of the NHSSM is given by%
\begin{eqnarray}
\label{NonH-TrilinearTerms}
-\cL_{\rm soft}^{\rm NH}&=&T_{ij}^{^\prime D}h_2 {\tilde d}_{Ri}^*{\tilde q}_{Lj}
+T_{ij}^{^\prime U}h_1 {\tilde u}_{Ri}^*{\tilde q}_{Lj}
+T_{ij}^{^\prime E}h_2 {\tilde e}_{Ri}^*{\tilde l}_{Lj}
+\mu^{\prime} {\tilde h}_1 {\tilde h}_2.
\end{eqnarray}
Here, $\mu^{\prime}$ represents the NH higgsino mass term, while $T_{ij}^{^\prime U}$, $T_{ij}^{^\prime D}$, and $T_{ij}^{^\prime E}$ are the NH trilinear couplings associated with up-type squarks, down-type squarks, and charged sleptons, respectively. The $h_1$ and $h_2$ denote the two Higgs doublets.  It is crucial to emphasize that these terms may not have a direct correlation with the holomorphic trilinear soft terms. In the presence of the NH trilinear terms, the charged slepton mass matrix can be written as,  
\begin{equation}
M_{\tilde{L}}^{2}=\left(
\begin{array}
[c]{cc}%
m_{\tilde{L}_{LL}}^{2} & m_{\tilde{L}_{LR}}^{2}\\[.5em]
m_{\tilde{L}_{LR}}^{2\dag} & m_{\tilde{E}_{RR}}^{2}%
\end{array}
\right) \label{fermion mass matrix}%
\end{equation}
with $m_{\tilde{L}_{LL}}^{2}$, $m_{\tilde{E}_{RR}}^{2}$ and $m_{\tilde{L}_{LR}}^{2}$ are the $3\times3$ matrices given by%
\begin{align}
m_{\tilde{L}_{LL_{ij}}}^{2}  & =m_{\tilde{L}_{ij}}^{2}+M_{Z}^{2}\cos2\beta\left(  I_{3}%
^{L}-Q_{L}s_{W}^{2}\right) \delta_{ij}  +m_{L}^{2}\delta_{ij},\nonumber\\
m_{\tilde{E}_{RR_{ij}}}^{2}  & =m_{\tilde{E}_{ij}}^{2}+M_{Z}^{2}\cos2\beta Q_{L}s_{W}^{2}\delta_{ij}+m_{L}^{2}\delta_{ij},\nonumber\\
m_{\tilde{L}_{LR_{ij}}}^{2}   & = \frac{v_1}{\sqrt{2}} \left(T_{ij}^{E}-(\mu Y_{ij}^{E}+T_{ij}^{^\prime E}) \tb\right).
\label{Eq:MSlepLR}
\end{align}
Here, $I_{3}^{L}$ represents the weak isospin of leptons, $Q_{L}$ stands for the electromagnetic charge, and $m_{L}$ denotes the mass of the SM leptons. The subscripts $L$ and $E$ refer to left- and right-handed sleptons, respectively. $\MZ$ and $\MW$ denote the masses of the $Z$ and $W$ bosons, while $s_W$ is defined as the square root of $1 - c_W^2$, where $c_W = \MW/\MZ$. Furthermore, $Y_{ij}^{E}$  is the Yukawa coupling associated with leptons, the $T_{ij}^{E}$ represent holomorphic trilinear couplings  and $\tb$ is defined as the ratio of the two vacuum expectation values of the two Higgs doublets, denoted by $v_1$ and $v_2$, $\tb=v_2/v_1$. It should be noted that because of the different combination of fields in $\cL_{\rm soft}^{\rm NH}$ w.r.t.\ the holomorphic SSB terms, the NH trilinear couplings $T_{ij}^{^\prime E}$ receive the additional factors of $\tb$ as can be seen in \refeq{Eq:MSlepLR}.

In addition to modifying the slepton mass matrix, the NH trilinear terms also modify the Higgs bosons coupling to charged sleptons. The couplings of the lightest Higgs boson $h$ to charged sleptons is given by,
\begin{align} 
C(h,\tilde{l}_{s},\tilde{l}_{t}) =
 &-\frac{i}{4} \sum_{i,j=1}^{3}\Big[ \Big(v_1 s_{\alpha}  + v_2 c_{\alpha} \Big) \Big( g_{2}^{2}R^{E,*}_{i s} R_{{i t}}^{E}-g_{1}^{2}  \Big( R^{E,*}_{i s} R_{{i t}}^{E} +R^{E,*}_{ i+3,s} R_{{i+3, t }}^{E}\Big)  \Big) \nonumber \\
&-\sum_{k=1}^{3} 4 v_1 s_{\alpha}  \Big( R^{E,*}_{k+3,s} Y_{ik}^{E} Y_{ij}^{E}  R_{{j+3,t}}^{E} -R^{E,*}_{js} Y_{ik}^{E} Y_{ij}^{E}  R_{{kt}}^{E} \Big)  \nonumber \\ 
&+ \sqrt{2} R^{E,*}_{js} R_{{i+3,t}}^{E} \Big(- 2 s_{\alpha}  T_{ij}^{E}  + c_{\alpha} \mu^* Y_{ij}^{E} + c_{\alpha} T_{ij}^{^\prime E}  \Big) \nonumber \\ 
 &+ \sqrt{2} R_{{jt}}^{E} R^{E,*}_{i+3,s} \Big( - 2 s_{\alpha}  T_{ij}^{E} +c_{\alpha} T_{ij}^{^\prime E} + c_{\alpha} \mu Y_{ij}^{E} \Big)\Big].
 \label{Eq:Higg-sferion-Coupling}
 \end{align} 
Here, $R^{E}_{st}$ denotes the $6\times6$ mixing matrix for charged sleptons, where the subscripts $s,t=1,2,..6$. The $g_{1}$ and $g_{2}$ stand for the gauge couplings corresponding to the $U(1)$ and $SU(2)$ gauge groups, respectively, while $c_{\alpha}$ ($s_{\alpha}$) represent the cosine (sine) of the mixing angle $\alpha$ that diagonalizes the $\cp$-even Higgs sector at tree-level.
\refeq{Eq:Higg-sferion-Coupling} shows that the NH terms play a direct role in the couplings of the Higgs bosons to sfermions. This insight is particularly crucial, especially concerning Higgs decays, and can potentially result in increased decay rates compared to holomorphic terms, given that the NH terms do not directly contribute to cLFV decays but only through the charged slepton mass matrix. 


\section{Calculation of low-energy observables}
\label{sec:CalcSetup}

\subsection{The cLFV decays}

\begin{figure}[htb!]
\centering
\includegraphics[width=12cm,height=8cm,keepaspectratio]{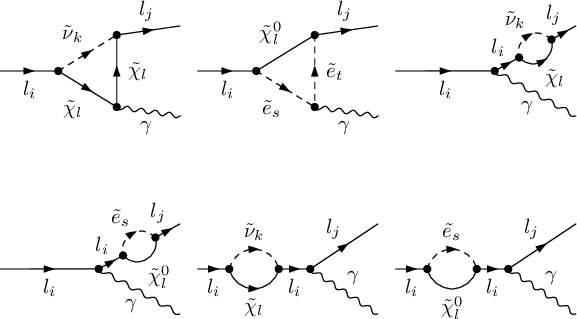}
\caption{Feynman diagrams for the NHSSM contributions to $l_{i} \rightarrow l_{j} \gamma$ with $i,j=1,2,3$ and $i>j$ (see text). }   
\label{fig:cLFV-FeynDiag}
\end{figure}

The branching ratios for the cLFV decays such as $l_{i} \rightarrow l_{j} \gamma$, where $i,j=1,2,3$ denote generations and $i>j$, can be significantly enhanced in SUSY models due to the presence of slepton flavor mixing\cite{Arana-Catania:2013nha,Gomez:2014uha}. In the NHSSM, Feynman diagrams depicting cLFV decays are shown in \reffi{fig:cLFV-FeynDiag}. Despite having the same Feynman diagrams as in the MSSM, the presence of NH terms in the charged lepton mass matrix, as shown in \refeq{Eq:MSlepLR}, and the Higgs sfermion couplings, as indicated in \refeq{Eq:Higg-sferion-Coupling}, can significantly alter the contribution of each diagram.  Within the NHSSM framework, diagrams involving sneutrinos $({\tilde \nu}_i)$ and charginos $({\tilde \chi}_l)$ in the loops generally have negligible impact\cite{Chattopadhyay:2019ycs}. However, those involving charged sleptons $({\tilde e}_s)$ and neutralinos $({\tilde \chi}_i^0)$ can be very significant, as the LFV NH terms contribute to the charged slepton mass matrix  enhanced by factor of $\tb$. Consequently, it is expected that the NH contributions will be dominant compared to the holomorphic ones unless $T_{ij}^{E}$ is very large or if substantial cancellation occur between holomorphic and NH contributions. 

\subsection{\boldmath{$h \to l_{i} l_{j}$} decays}

\begin{figure}[htb!]
\centering
\includegraphics[width=15cm,height=10cm,keepaspectratio]{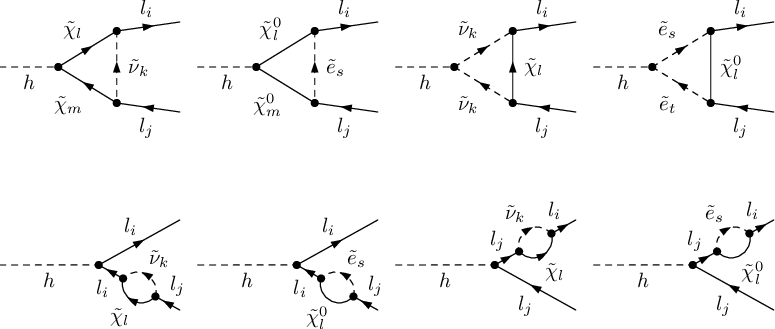}
\caption{Feynman diagrams for the NHSSM contributions to $h \rightarrow l_{i} l_{j}$ with $i,j=1,2,3$ and $i\neq j$ (see text).} 
\label{fig:LFVHD-FeynDiag}
\end{figure}

The Feynman diagrams for $h \rightarrow l_{i} l_{j}$ with $i\neq j$ are shown in \reffi{fig:LFVHD-FeynDiag}. As discussed in the preceding section, the diagrams involving sneutrinos $({\tilde \nu}_i)$ and charginos $({\tilde \chi}_l)$ remain unaffected by the NH terms for the case of LFVHD as well. However, those involving charged sleptons $({\tilde e}_s)$ and neutralinos $({\tilde \chi}_i^0)$ in the loop can have a significant impact. Moreover, in addition to being influenced by the LFV originating from the charged slepton mass matrix\cite{Gomez:2017dhl}, the LFVHD are further affected by the presence of LFV NH terms in the coupling of the Higgs boson to charged sleptons. 

\subsection{Computational Setup}
\label{sec:comp-setup}
For our analysis, we employed the Mathematica package {\tt SARAH}~\cite{Staub:2009bi,Staub:2010jh,Staub:2012pb,Staub:2013tta,Staub:2015kfa} to generate the source code for {\tt SPheno}~\cite{Porod:2003um}. This source code encompasses analytical expressions pertinent to NHSSM, including mass matrices, couplings, decay rates, and branching ratios. The input parameters are provided to {\tt SPheno} in the form of an SLHA~\cite{Allanach:2008qq} file. By utilizing this {\tt SPheno} source code, via the corresponding SLHA output, one can obtain numerical results for mass spectra and other desired observables, such as cLFV decays and LFVHD, for any specified set of input parameters. 

The {\tt SARAH Scan and Plot} ({\tt SSP}) \cite{Staub:2011dp} package is then employed to input parameters within a specified range. For this purpose, the {\tt SSP} package generates an SLHA file and invokes {\tt SPheno} within Mathematica. The output produced by {\tt SPheno} is subsequently read by {\tt SSP} and stored in a Mathematica-readable format. By utilizing {\tt SSP}, one can also apply various constraints to the {\tt SPheno}-generated output to restrict parameters to within the allowed range. We employ two types of constraints: the cLFV constraints (detailed further in \refse{sec:cLFV-Constraints}) and the charge and color-breaking global minima (CCB) constraints.

It is known from the literature\cite{Beuria:2017gtf} that a substantial trilinear coupling, holomorphic or NH, generally leads to non-physical or metastable CCB minima. In the context of LFV, this concerns primarily a charge breaking vacuum, resulting from the involvement of off-diagonal entries of the trilinear couplings. The CCB constraints in the context of NHSSM were studied in detail in a ~\citere{Chattopadhyay:2019ycs}. To incorporate these constraints in our analysis, we developed a private Mathematica code using the expressions given in~\citere{Chattopadhyay:2019ycs}. However, as we will see in the following section, the CCB constraints had little to no impact on our results.

\section{Numerical Results}
\label{sec:NResults}

\subsection{Input parameters}
\label{sec:input-para}

For our numerical analysis, we have selected three scenarios denoted as
$M_{h}^{125}$, $M_{h}^{125}(\tilde{\tau})$, and
$M_{h}^{125}(\tilde{\chi})$, originally introduced in
\citere{Bahl:2018zmf}. These scenarios are parameterized in terms of
electroweak-scale parameters, carefully chosen to highlight various
aspects of Higgs boson phenomenology within the MSSM. It is noteworthy
that these scenarios, across a broad range of their parameter space, are
in agreement with the experimental findings from the LHC regarding the
properties of the Higgs boson, as well as constraints on masses and
couplings of new particles. Each scenario features a $\cp$-even scalar
with a mass approximately around 125 GeV, exhibiting SM-like
couplings (see also \citere{Slavich:2020zjv}).
The first scenario, $M_{h}^{125}$, is characterized by
relatively heavy superparticles, leading to Higgs phenomenology at the
LHC resembling that of a Two-Higgs-Doublet Model with Higgs couplings
inspired by the MSSM. On the other hand, the second and third scenarios,
$M_{h}^{125}(\tilde{\tau})$ and $M_{h}^{125}(\tilde{\chi})$, involve
some of the superparticles, namely staus and charginos and neutralinos
respectively, being relatively light. This has implications for the
decays of the heavier Higgs bosons, resulting in weakened exclusion
bounds from searches involving $\tau^{+}\tau^{-}$, as well as affecting
the decay of the lighter $\cp$-even scalar to photons. Importantly,
these scenarios open up the possibility of exploring additional Higgs
bosons through their decays to charginos and neutralinos. 

Encompassing a diverse range of MSSM Higgs-boson phenomenology, the
benchmark scenarios shown in
~\refta{tab:input-parameters}~\cite{Bahl:2018zmf} can serve as 
guidelines and motivation for ongoing/upcoming LHC searches targeting
additional neutral and charged Higgs bosons. 

\begin{table}[htbp] 
\centerline{\begin{tabular}{|c|c|c|c|}
\hline\hline
& $M_{h}^{125}$  & $M_{h}^{125}(\tilde{\tau})$ & $M_{h}^{125}%
(\tilde{\chi})$ \\\hline\hline
$m_{\tilde{Q}_{3}},m_{\tilde{U}_{3}},m_{\tilde{D}_{3}}$ & $1500$ & $1500$ &
$1500$\\
$m_{\tilde{L}_{3}},m_{\tilde{E}_{3}}$ & $2000$ & $350$ & $2000$\\
$\mu$ & $1000$ & $1000$ & $180$\\
$M_{1}$ & $1000$ & $180$ & $160$\\
$M_{2}$ & $1000$ & $300$ & $180$\\
$M_{3}$ & $2500$ & $2500$ & $2500$\\
$X_{t}$ & $2800$ & $2800$ & $2500$\\
$A_{\tau}$ & $0$ & $800$ & $0$\\
\hline\hline
\end{tabular}}
\caption{Selected scenarios in the MSSM parameter space, taken from
  \citere{Bahl:2018zmf}. All the dimensionful quantities are in $\gev$.}%
\label{tab:input-parameters}
\end{table}%

In \refta{tab:input-parameters}, $m_{\tilde{Q}_{3}}$,
$m_{\tilde{U}_{3}}$, and $m_{\tilde{D}_{3}}$ denote the third generation
masses of the squark doublet, up-type squark singlet, and down-type
squark singlet, respectively. Additionally, $m_{\tilde{L}_{3}}$ and
$m_{\tilde{E}_{3}}$ denote the third generation masses of the
left-handed slepton doublet and right-handed slepton singlet
respectively. The $M_{1}$, $M_{2}$ and $M_{3}$ are the gaugino masses
and $\mu$ is the usual Higgs mixing parameter. The
$X_t= A_t-\mu \cot \beta$ with $A_t=T_t/Y_t$ and $A_\tau=T_\tau/ Y_\tau$ are the 
holomorphic trilinear couplings.   In these benchmark scenarios, the SSB 
mass $M_{\tilde f}$ for the first two generations is chosen to be $2 
\tev$, and the holomorphic trilinear SSB terms for these generations are 
taken to be zero. For each scenario, we investigate four different 
combinations of $M_A$ and $\tb$, taking into account the latest
experimental limits for MSSM 
Higgs-boson searches~\cite{ATLAS:2020zms,CMS:2022goy,ATLAS:2024lyh}: 
\begin{align*}
\rm P1 &:~\MA = 1500 \gev,~ \tb = 7 \\
\rm P2 &:~\MA = 2000 \gev,~ \tb = 15 \\
\rm P3 &:~\MA = 2500 \gev,~ \tb = 30 \\
\rm P4 &:~\MA = 2500 \gev,~ \tb = 45 
\end{align*}

In the definition of the benchmark scenarios several indirect
constraints such as dark matter relic density and the anomalous magnetic
moment of muon were not taken into account, as their impact on
parameters associated with Higgs phenomenology is minimal.  

For our numerical analyses, for each scenario mentioned in
\refta{tab:input-parameters}, and for different combinations of $M_A$
and $\tb$ as specified above, we vary one off-diagonal entry of either
$T_{ij}^{E}$ or $T_{ij}^{\prime E}$ within the range of
$(-10000:10000)\gev$, while setting all other off-diagonal entries to
zero. Subsequently, we only consider those values of $T_{ij}^{E}$ and
$T_{ij}^{\prime E}$ that comply with the constraints imposed by cLFV
decays, in order to obtain predictions for LFVHD.
Here, we indicate the regions of parameter space that are already
probed by the ATLAS and CMS experiments, particularly for $h \to e \tau$
and $h \to \mu\tau$.


\subsection{cLFV Constraints}
\label{sec:cLFV-Constraints}

Decays involving cLFV, such as $l_{i} \to l_{j} \gamma$, where
$l_{i}$ represents a charged lepton with the generation index
$i,j=1,2,3$ and $i>j$, impose the most rigorous constraints on LFV due
to exceptionally tight limits on these decay processes. Notably, the
upper limits on the processes involving $\mu-e$ transition such as $\mu
\to e \gamma$ and conversion rate ${\rm CR}(\mu-e, {\rm Au})$,
are particularly stringent, significantly constraining LFV within the
$\mu-e$ sector. In contrast, the constraints on transitions involving
$\tau-e$ and $\tau-\mu$ are comparatively weaker, allowing for
considerable LFV that can be explored in present experiments. 

In \refta{tab:cLFV-BR-Limits}, we list the present constraints on
cLFV~\cite{Workman:2022ynf} and LFVHD decays. Concerning the
latter, the limits on decays involving taus are based on 
\citere{CMS:2021rsq} (later labeled as ``CMS''), whereas 
the $h \to e \mu$ limit is taken from \citere{ATLAS:2019old}.

\begin{table}[htbp]
\centerline{\begin{tabular}{|c|c||c|c|}
\hline\hline
 Process &  Limit &  Process &  Limit   \\ 
 \hline\hline
$\br(\mu \to e \gamma)$ & $4.2 \times 10^{-13}$ & $\br(\mu \to 3e)$ &   $1.0 \times 10^{-12}$  \\ 
${\rm CR}(\mu-e, {\rm Au})$ &  $7.0 \times 10^{-13}$ & ${\rm CR}(\mu-e, {\rm Ti})$ & $4.3 \times 10^{-13}$  \\ 
$\br(\tau \to e \gamma)$ & $3.3\times 10^{-8}$ & $\br(\tau \to 3e)$ & $2.7\times 10^{-8}$  \\ 
$\br(\tau \to \mu \gamma)$ & $4.4 \times 10^{-8}$ & $\br(\tau \to 3\mu)$ & $2.1 \times 10^{-8}$  \\ \hline
$\br(h \to e \mu)$ & $6.1 \times 10^{-5}$ & $\br(h \to e \tau)$ & $2.2 \times 10^{-3}$  \\ 
$\br(h \to \mu \tau)$ & $1.5 \times 10^{-3}$ &  &   \\ 
\hline\hline
\end{tabular}}
\caption{Present upper bounds on the cLFV decays and LFV Higgs
  decays~\cite{Workman:2022ynf,CMS:2021rsq,ATLAS:2019old} (see text).
  The bounds on LFVHD involving
  taus are indicated in the plots where applicable (labeled as ``CMS''). }
\label{tab:cLFV-BR-Limits}
\end{table}


The limits presented in Tab.~\ref{tab:cLFV-BR-Limits} impose stringent
constraints on the LFV holomorphic and NH SSB terms. By utilizing the
input parameters provided in \refta{tab:input-parameters}, we have
computed the allowed range for $T_{ij}^{E}$ and $T_{ij}^{^\prime E}$. It
is assumed that only one of the couplings $T_{ij}^{E}$ or
$T_{ij}^{\prime E}$ is non-zero at a given time, where $T_{ij} = Y_{ij}
A_{ij}$ and $T_{ij}^{\prime} = Y_{ij} A_{ij}^{\prime}$. A non-zero value
of either $T_{ij}^{E}$ or $T_{ij}^{\prime E}$ will result in a non-zero
value of $Y_{ij}^{E}$. In \refta{tab:cLFV-Limits}, we show the
constraints on $T_{ij}^{E}$ and $T_{ij}^{^\prime E}$ in the
$M_{h}^{125}$, $M_{h}^{125}(\tilde{\tau})$, and
$M_{h}^{125}(\tilde{\chi})$ scenarios.

\renewcommand{\arraystretch}{1.2}
\begin{table}[h!]
\centerline{\begin{tabular}{|c|c|c|c|c|c|c|c|c|c|c|c|c|}
\hline\hline
\multicolumn{13}{|c|}{$M_{h}^{125}$} \\
\hline\hline
  & $|T_{12}^{E}|$ &  $|T_{12}^{^\prime E}|$ &  $|T_{21}^{E}|$ &  $|T_{21}^{^\prime E}|$ & $T_{13}^{E}$ & $|T_{13}^{^\prime E}|$ & $|T_{31}^{E}|$ & $|T_{31}^{^\prime E}|$ & $|T_{23}^{E}|$ & $|T_{23}^{^\prime E}|$ & $|T_{32}^{E}|$ & $|T_{32}^{^\prime E}|$  \\
\hline 
$P1 $  & 0.946 &  0.135 &  0.946 &  0.135 & 4855 & 1486 & 5108 & 1432 & 5108 & 1702 & 5324 & 1630 \\
$P2 $  & 2.027 &  0.135 &  2.027 &  0.135 & 5972 & 1486 & 6441 & 1432 & 6153 & 1684 & 6549 & 1630 \\
$P3 $  & 4.099 &  0.135 &  4.099 &  0.135 & 6549 & 1468 & 7095 & 1414 & 6675 & 1684 & 7095 & 1612 \\
$P4 $  & 6.189 &  0.135 &  6.189 &  0.135 & 6747 & 1468 & 7109 & 1396 & 6855 & 1666 & 7109 & 1612 \\
\hline
\multicolumn{13}{|c|}{$M_{h}^{125}(\tilde{\tau})$} \\
\hline\hline
  & $|T_{12}^{E}|$ &  $|T_{12}^{^\prime E}|$ &  $|T_{21}^{E}|$ &  $|T_{21}^{^\prime E}|$ & $T_{13}^{E}$ & $|T_{13}^{^\prime E}|$ & $|T_{31}^{E}|$ & $|T_{31}^{^\prime E}|$ & $|T_{23}^{E}|$ & $|T_{23}^{^\prime E}|$ & $|T_{32}^{E}|$ & $|T_{32}^{^\prime E}|$  \\
\hline 
$P1 $  & 2.460 &  0.375 &  2.460 &  0.375 & 1006 & 198.0 & 909.0 & 209.0 & 1096 & 228.0 & 963.0 & 240.0 \\
$P2 $  & 5.270 &  0.373 &  5.270 &  0.375 & 1406 & 187.0 & 1180 & 208.0 & 1468 & 215.0 & 1216 & 239.0 \\
$P3 $  & 10.54 &  0.373 &  10.59 &  0.373 & 1396 & 149.0 & 1085 & 169.0 & 1414 & 172.0 & 1085 & 194.0 \\
\hline
\multicolumn{13}{|c|}{$M_{h}^{125}(\tilde{\chi})$} \\
\hline\hline
  & $|T_{12}^{E}|$ &  $|T_{12}^{^\prime E}|$ &  $|T_{21}^{E}|$ &  $|T_{21}^{^\prime E}|$ & $T_{13}^{E}$ & $|T_{13}^{^\prime E}|$ & $|T_{31}^{E}|$ & $|T_{31}^{^\prime E}|$ & $|T_{23}^{E}|$ & $|T_{23}^{^\prime E}|$ & $|T_{32}^{E}|$ & $|T_{32}^{^\prime E}|$  \\
\hline 
$P1 $  & 2.710 &  0.405 &  2.710 &  0.405 & 4846 & 3234 & 6135 & 3108 & 5972 & 3504 & 6225 & 3378 \\
$P2 $  & 5.790 &  0.405 &  5.790 &  0.405 & 6405 & 3216 & 6783 & 3108 & 6477 & 3486 & 6801 & 3378 \\
$P3 $  & 11.62 &  0.405 &  11.62 &  0.405 & 6657 & 3216 & 4261 & 3072 & 6729 & 3486 & 4747 & 3342 \\
$P4 $  & 17.74 &  0.405 &  17.56 &  0.405 & 6783 & 3216 & 2972 & 3036 & 6837 & 3468 & 3432 & 3306 \\
\hline\hline
\end{tabular}}
\caption{Upper limits on $T_{ij}^{E}$ and $T_{ij}^{^\prime E}$ arising from cLFV decays within the $M_{h}^{125}$, $M_{h}^{125}(\tilde{\tau})$ and $M_{h}^{125}(\tilde{\chi})$ scenarios. All the dimensionful quantities are in $\gev$. }
\label{tab:cLFV-Limits}
\end{table}

As evident from \refta{tab:cLFV-Limits}, the couplings $T_{ij}^{ E}$ and
$T_{ij}^{^\prime E}$ involving the $\mu-e$ transition stand out as the
most constrained. This can be primarily attributed to the strong
experimental constraints on $\br(\mu \to e \gamma)$. In contrast,
the couplings related to $\tau-e$ or $\tau-\mu$ transitions can be
larger by $\order{10^{3}}$, mainly due to the comparatively weaker
experimental constraints on $\br(\tau \to e \gamma)$ and ${\rm
  BR}(\tau \to \mu \gamma)$, which are of $\order{10^{-8}}$. Despite the
fact that cLFV decays are the primary constraining factor, certain
values of $T_{ij}^{E}$ and $T_{ij}^{\prime E}$ are excluded because they
lead to negative scalar lepton masses. For instance, the point P4, where
$\tb=45$ in the $M_{h}^{125}(\tilde{\tau})$ scenario, results in
negative scalar lepton masses and is thus excluded.


\pagebreak
\subsection{\boldmath{$h \to l_{i} l_{j}$} decays} 

The large values of both holomorphic and NH trilinear couplings in the
$\tau-e$ and $\tau-\mu$ sectors can result in large branching ratios for
LFVHD. Specifically, the NH terms can have non-negligible effects on
Higgs phenomenology, as these terms directly contribute to the Higgs
coupling with the sleptons.

\begin{figure}
\centering
\includegraphics[width=7.3cm,height=5cm,keepaspectratio]{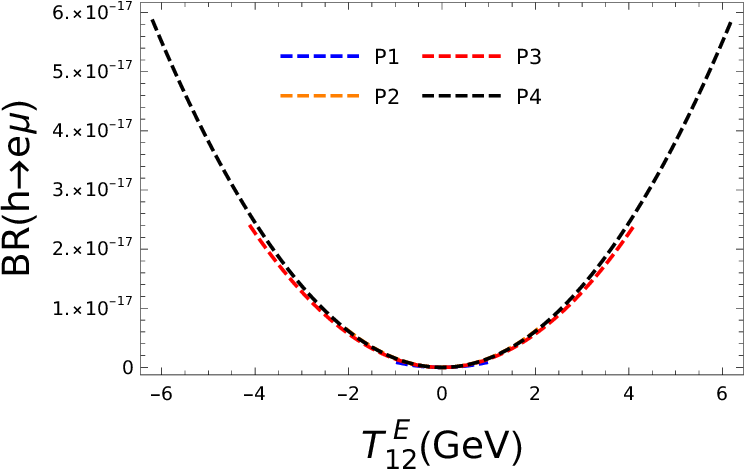}
\hspace{0.02\textwidth}
\includegraphics[width=7.3cm,height=5cm,keepaspectratio]{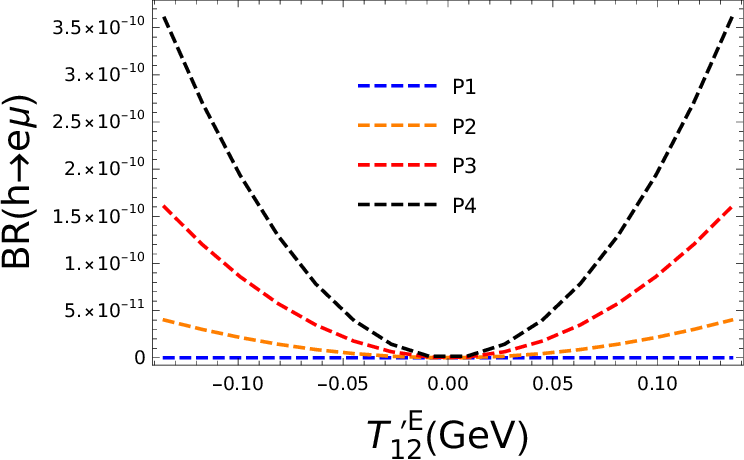}\\[0.5em]
\includegraphics[width=7.3cm,height=5cm,keepaspectratio]{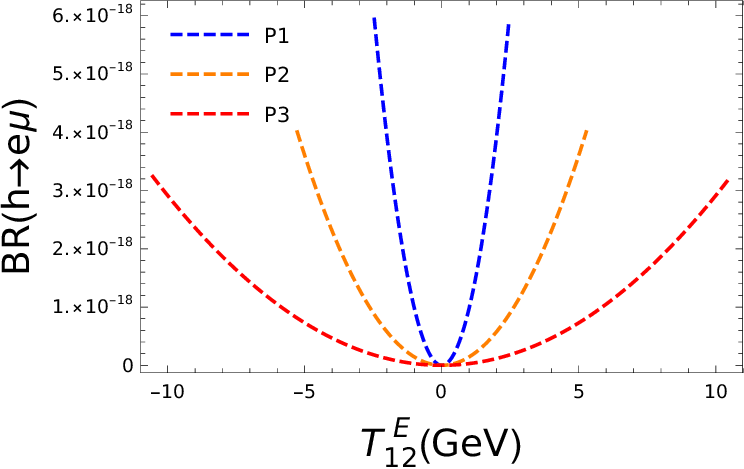}
\hspace{0.02\textwidth}
\includegraphics[width=7.3cm,height=5cm,keepaspectratio]{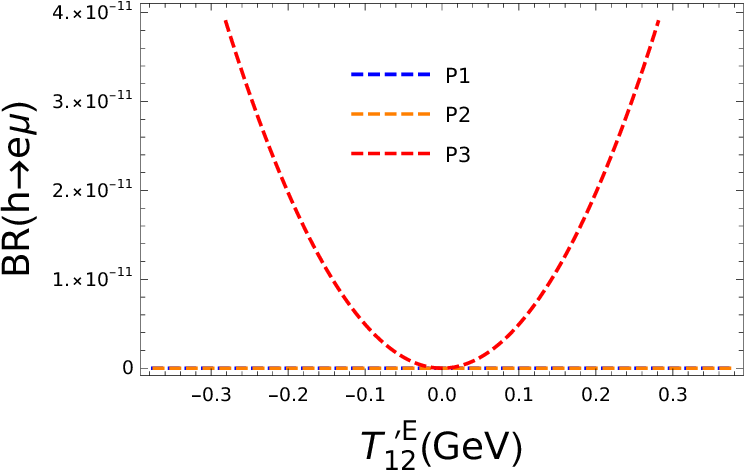}\\
\vspace{0.02\textwidth}
\includegraphics[width=7.3cm,height=5cm,keepaspectratio]{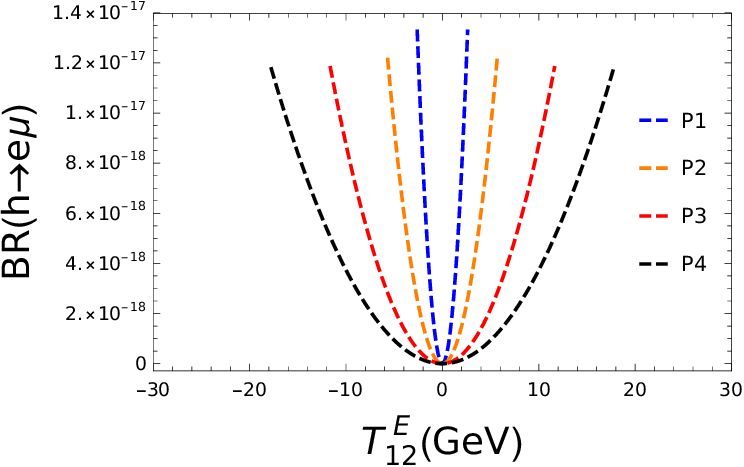}
\hspace{0.02\textwidth}
\includegraphics[width=7.3cm,height=5cm,keepaspectratio]{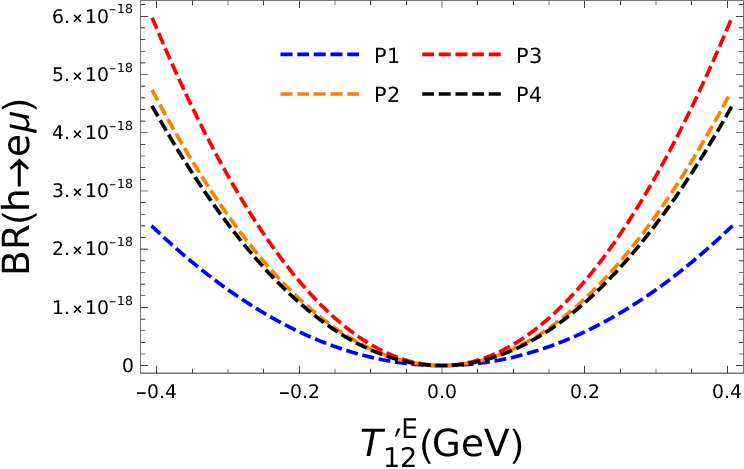}
\caption{\br($h \to e \mu)$ as a function of $T_{12}^{E}$ (left column) and $T_{12}^{\prime E}$ (right column) within the $M_{h}^{125}$ (upper row), $M_{h}^{125}(\tilde{\tau})$ (middle row) and $M_{h}^{125}(\tilde{\chi})$ (lower row) scenarios.}
\label{fig:Hemu-12}
\end{figure}

\begin{figure}
\centering
\includegraphics[width=7.0cm,height=5cm,keepaspectratio]{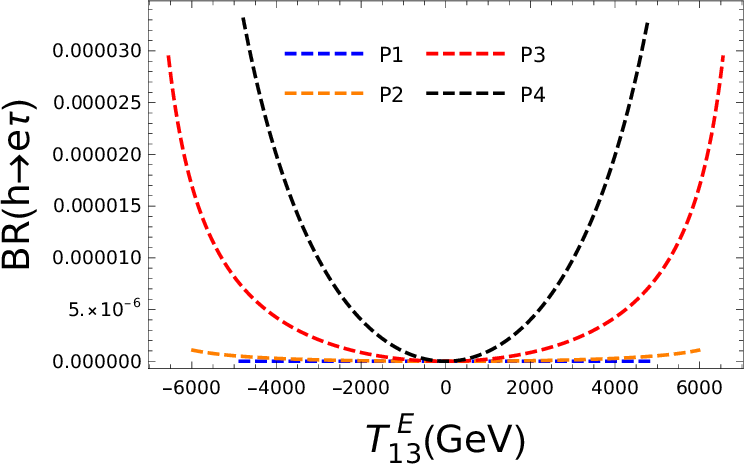}
\hspace{0.02\textwidth}
\includegraphics[width=7.0cm,height=5cm,keepaspectratio]{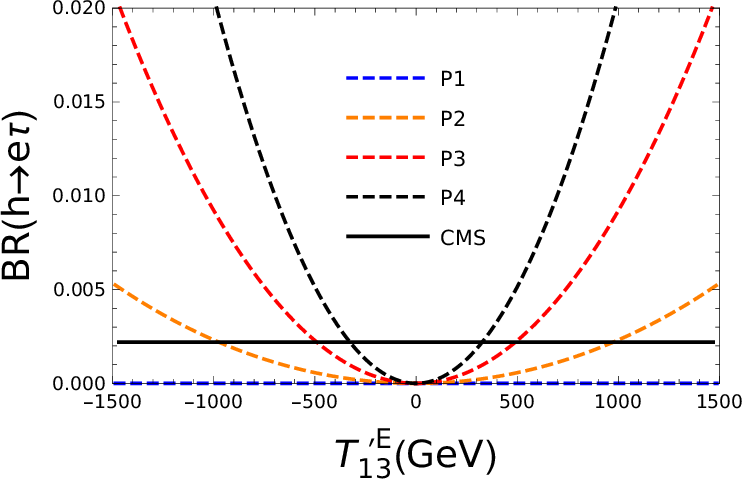}\\
\vspace{0.02\textwidth}
\includegraphics[width=7.0cm,height=4.3cm,keepaspectratio]{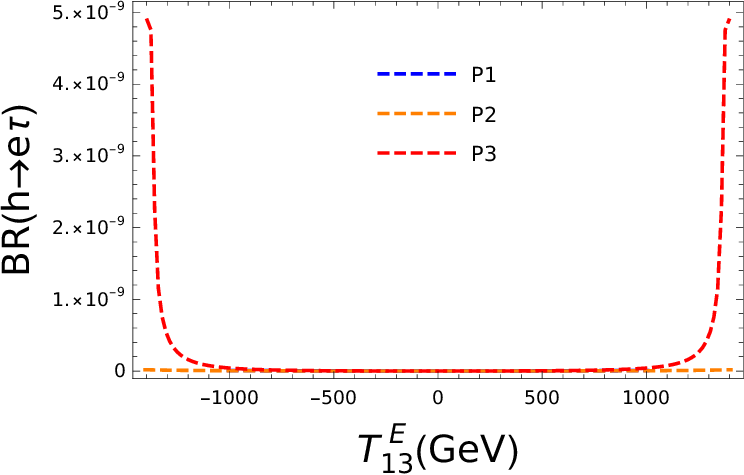}
\hspace{0.02\textwidth}
\includegraphics[width=7.0cm,height=5cm,keepaspectratio]{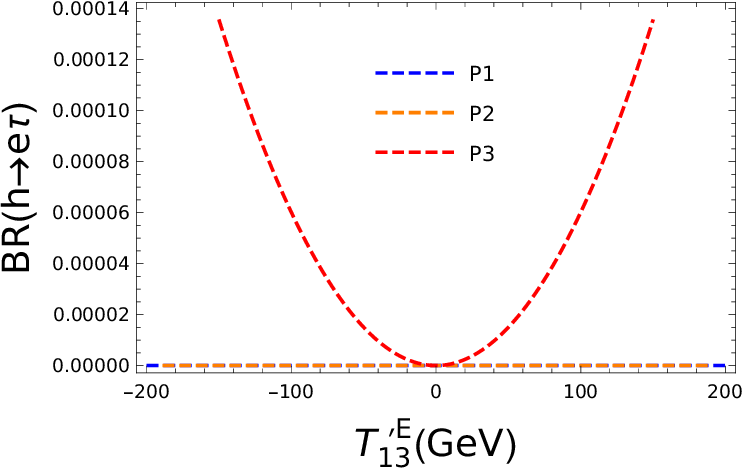}\\
\vspace{0.02\textwidth}
\includegraphics[width=7.0cm,height=5cm,keepaspectratio]{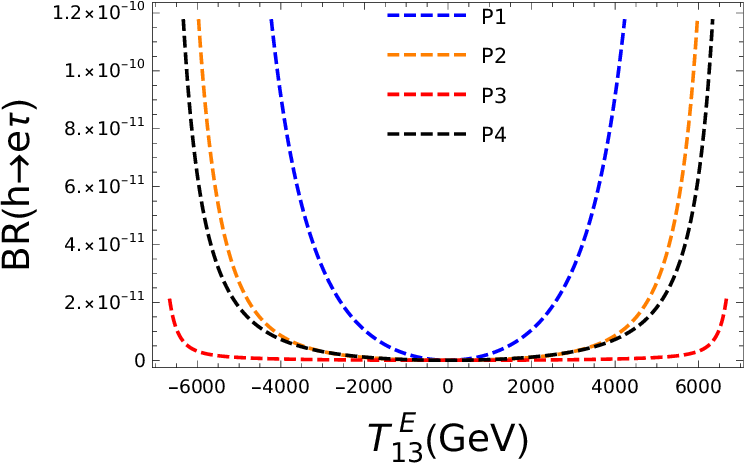}
\hspace{0.02\textwidth}
\includegraphics[width=7.0cm,height=5cm,keepaspectratio]{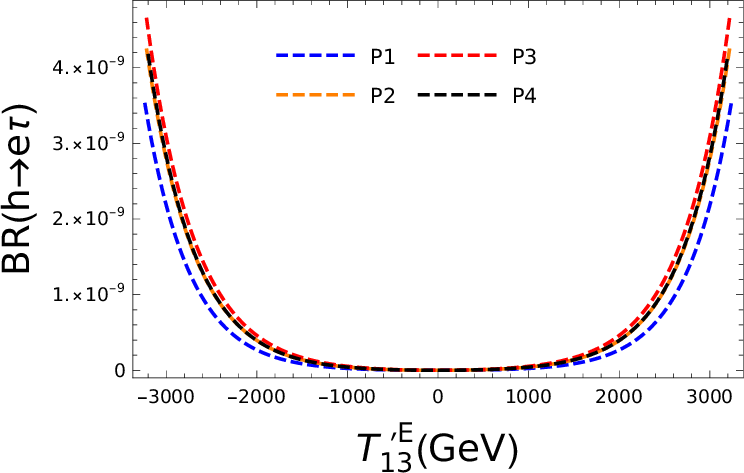}
\caption{$\br(h \to e \tau)$ as a function of $T_{13}^{E}$ (left column)
  and $T_{13}^{\prime E}$ (right column) within the $M_{h}^{125}$ (upper
  row), $M_{h}^{125}(\tilde{\tau})$ (middle row) and
  $M_{h}^{125}(\tilde{\chi})$ (lower row) scenarios. The horizontal
  solid line in the upper right plot indicates the experimental limit,
  see \protect\refta{tab:cLFV-BR-Limits}.}
\label{fig:Hetau-13}
\end{figure}

\begin{figure}[htb!]
\centering
\includegraphics[width=7.0cm,height=5cm,keepaspectratio]{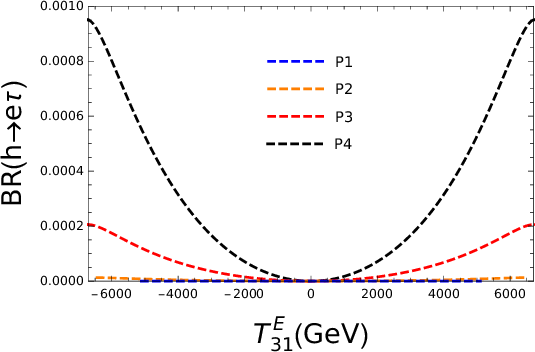}
\hspace{0.02\textwidth}
\includegraphics[width=7.0cm,height=5cm,keepaspectratio]{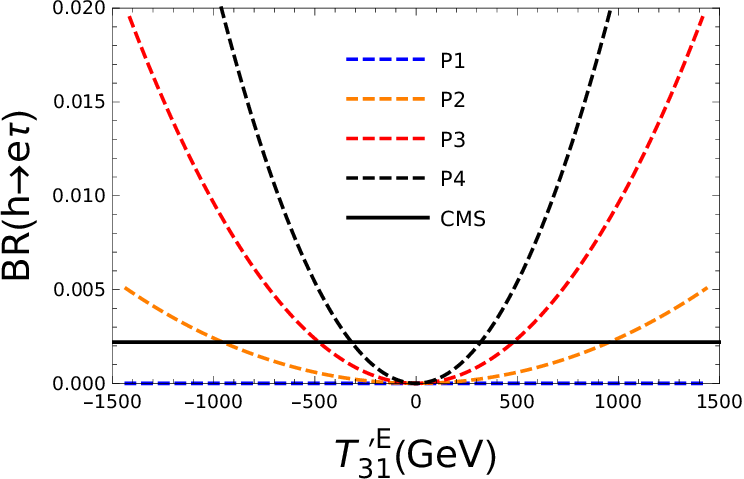}\\
\vspace{0.02\textwidth}
\includegraphics[width=7.0cm,height=4.3cm,keepaspectratio]{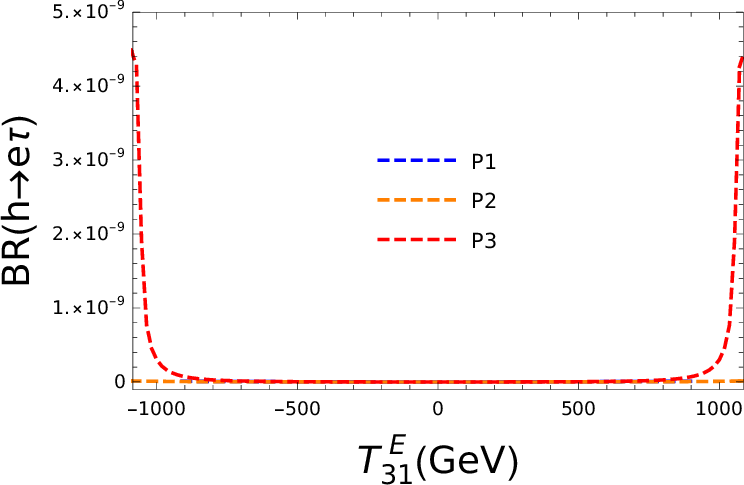}
\hspace{0.02\textwidth}
\includegraphics[width=7.0cm,height=5cm,keepaspectratio]{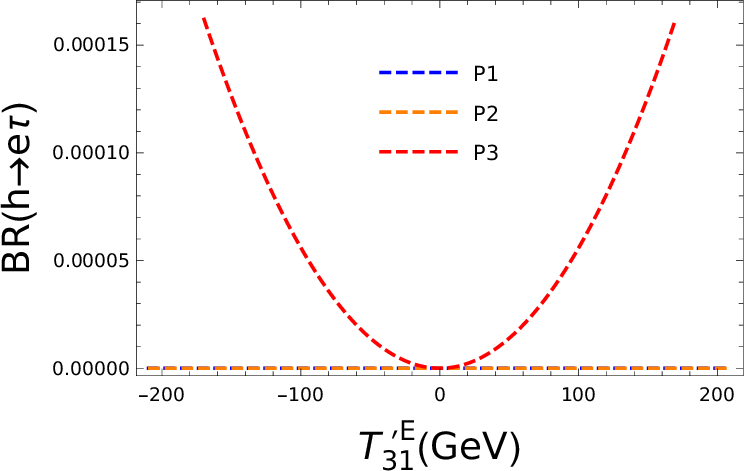}\\
\vspace{0.02\textwidth}
\includegraphics[width=7.0cm,height=5cm,keepaspectratio]{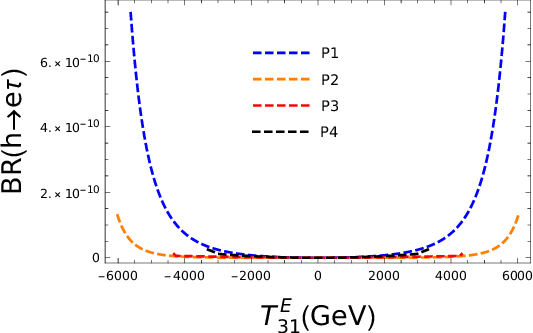}
\hspace{0.02\textwidth}
\includegraphics[width=7.0cm,height=5cm,keepaspectratio]{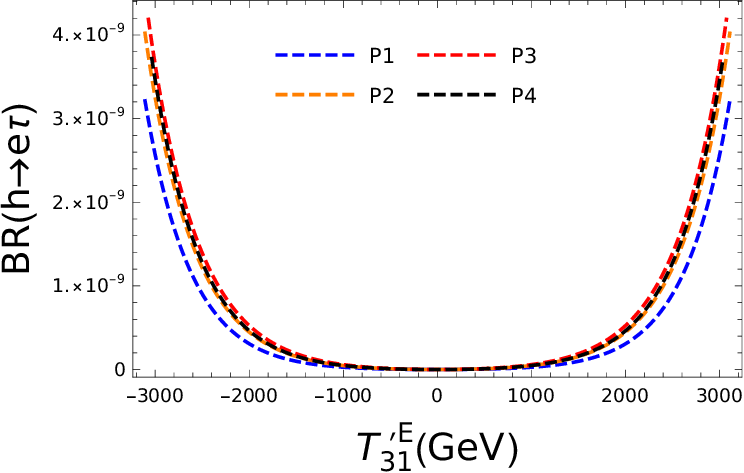}
\caption{$\br(h \to e \tau)$ as a function of $T_{31}^{E}$ (left
  column) and $T_{31}^{\prime E}$ (right column) within the
  $M_{h}^{125}$ (upper row), $M_{h}^{125}(\tilde{\tau})$ (middle row)
  and $M_{h}^{125}(\tilde{\chi})$ (lower row) scenarios.
The horizontal
  solid line in the upper right plot indicates the experimental limit,
  see \protect\refta{tab:cLFV-BR-Limits}.
}
\label{fig:Hetau-31}
\end{figure}

\begin{figure}[htb!]
\centering
\includegraphics[width=7.0cm,height=5cm,keepaspectratio]{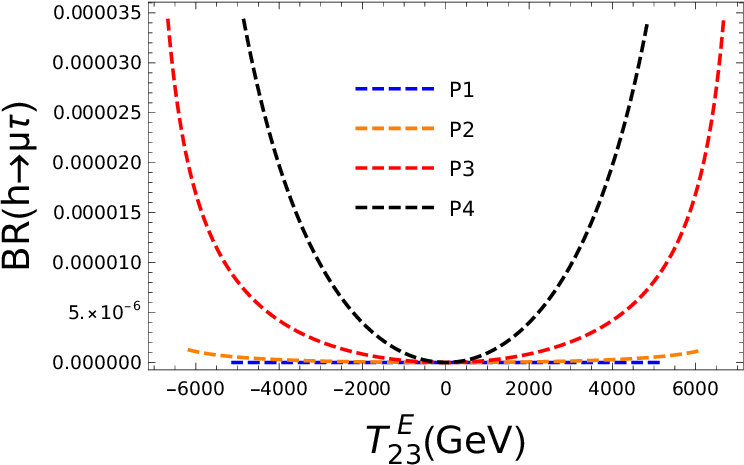}
\hspace{0.02\textwidth}
\includegraphics[width=7.0cm,height=4.3cm,keepaspectratio]{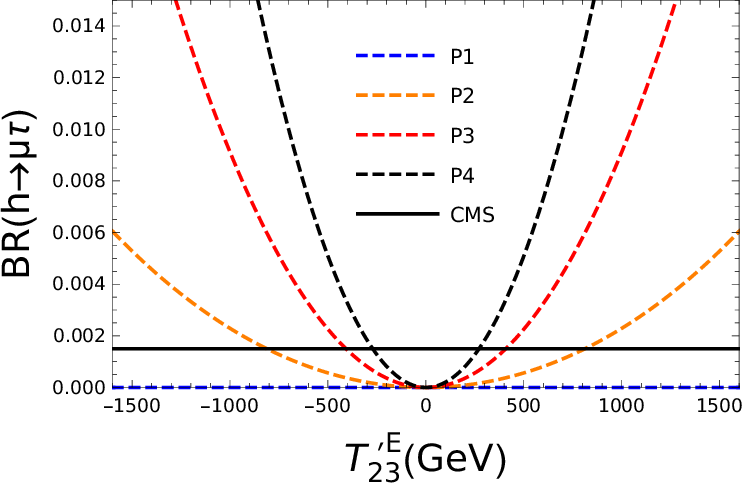}\\
\vspace{0.02\textwidth}
\includegraphics[width=7.0cm,height=4.3cm,keepaspectratio]{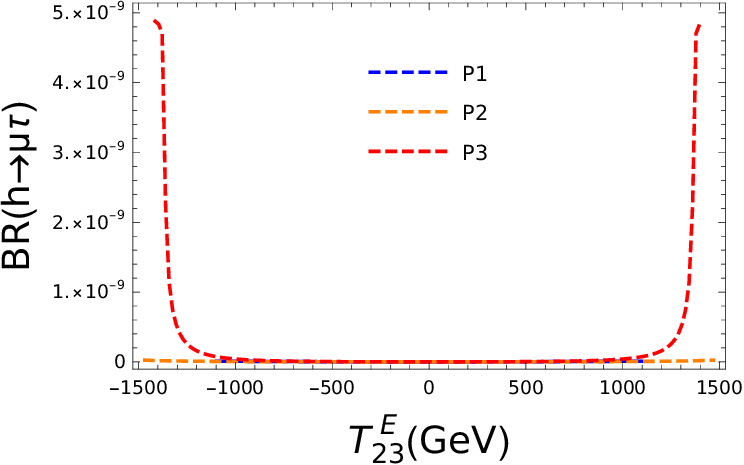}
\hspace{0.02\textwidth}
\includegraphics[width=7.0cm,height=5cm,keepaspectratio]{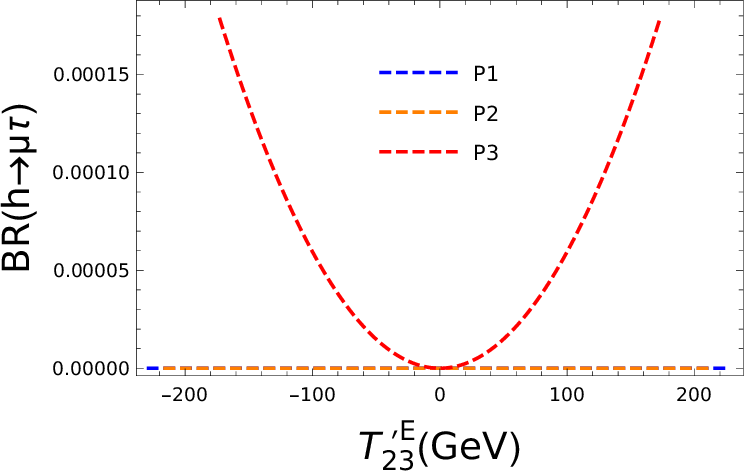}\\
\vspace{0.02\textwidth}
\includegraphics[width=7.0cm,height=5cm,keepaspectratio]{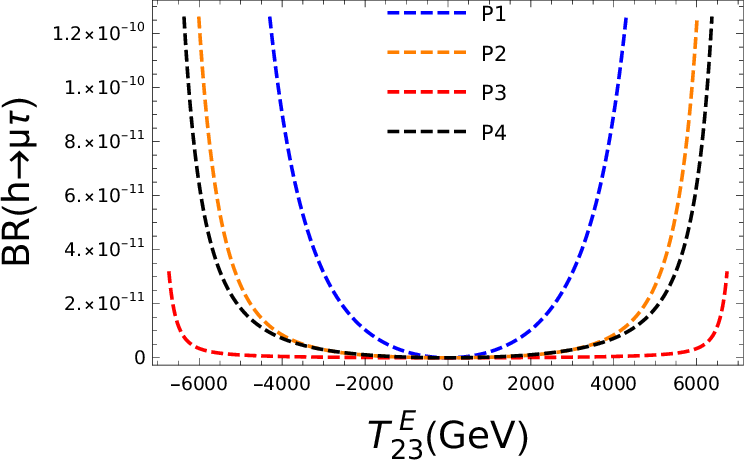}
\hspace{0.02\textwidth}
\includegraphics[width=7.0cm,height=4.3cm,keepaspectratio]{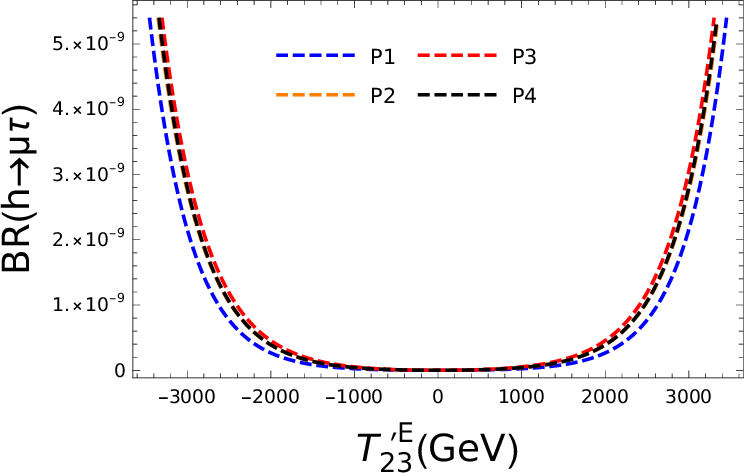}
\caption{$\br(h \to \mu \tau)$ as a function of $T_{23}^{E}$ (left
  column) and $T_{23}^{\prime E}$ (right column) within the
  $M_{h}^{125}$ (upper row), $M_{h}^{125}(\tilde{\tau})$ (middle row)
  and $M_{h}^{125}(\tilde{\chi})$ (lower row) scenarios.
The horizontal
  solid line in the upper right plot indicates the experimental limit,
  see \protect\refta{tab:cLFV-BR-Limits}.}
\label{fig:Hmutau-23}
\end{figure}

\begin{figure}[htb!]
\centering
\includegraphics[width=7.0cm,height=4.4cm,keepaspectratio]{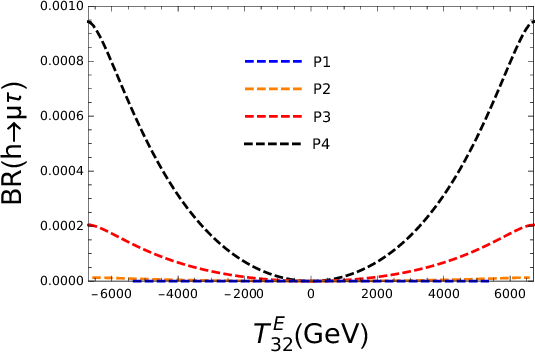}
\hspace{0.02\textwidth}
\includegraphics[width=7.0cm,height=4.3cm,keepaspectratio]{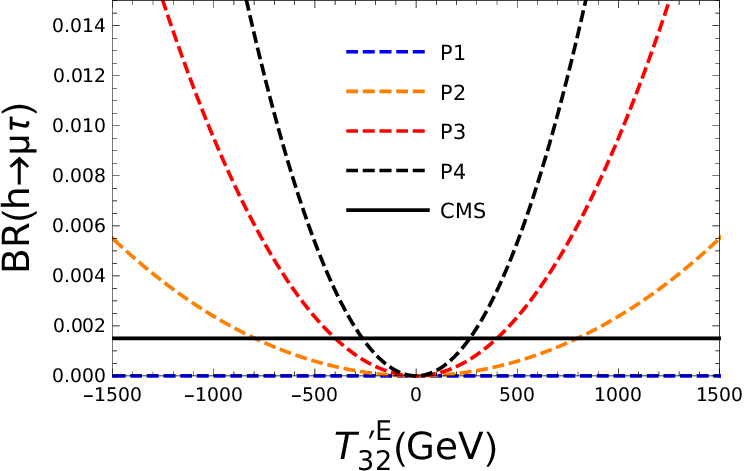}\\
\vspace{0.02\textwidth}
\includegraphics[width=7.0cm,height=4.3cm,keepaspectratio]{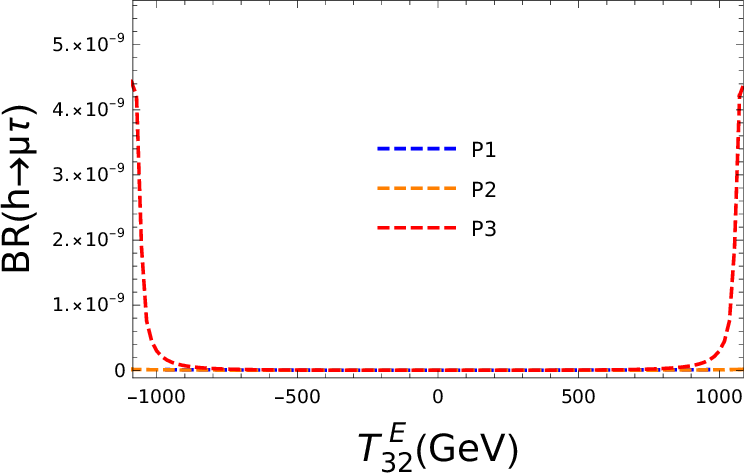}
\hspace{0.02\textwidth}
\includegraphics[width=7.0cm,height=5cm,keepaspectratio]{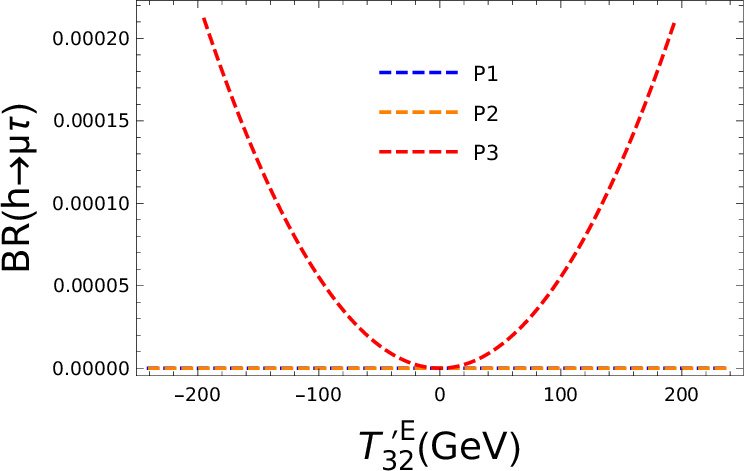}\\
\vspace{0.02\textwidth}
\includegraphics[width=7.0cm,height=5cm,keepaspectratio]{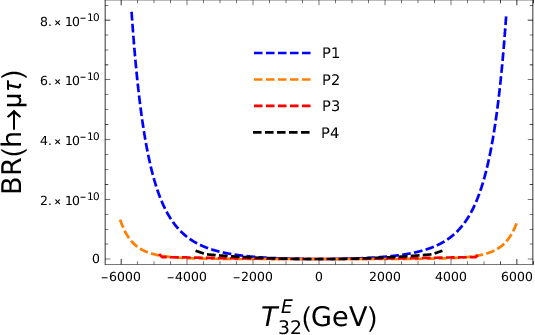}
\hspace{0.02\textwidth}
\includegraphics[width=7.0cm,height=4.3cm,keepaspectratio]{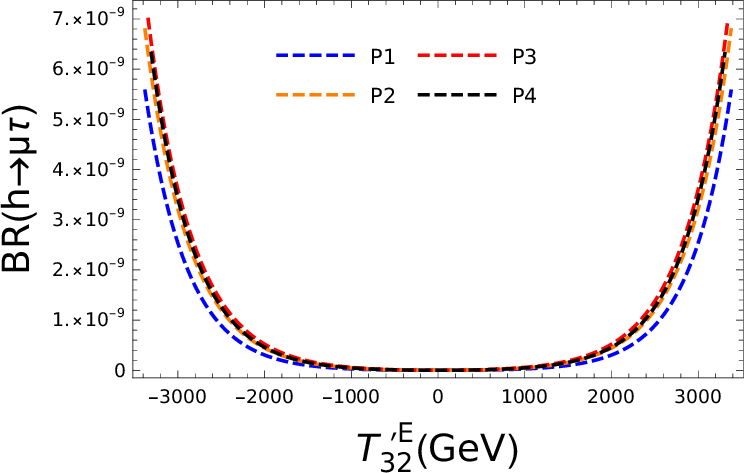}
\caption{$\br(h \to \mu \tau)$ as a function of $T_{32}^{E}$ (left
  column) and $T_{32}^{\prime E}$ (right column) within the
  $M_{h}^{125}$ (upper row), $M_{h}^{125}(\tilde{\tau})$ (middle row)
  and $M_{h}^{125}(\tilde{\chi})$ (lower row) scenarios.
The horizontal
  solid line in the upper right plot indicates the experimental limit,
  see \protect\refta{tab:cLFV-BR-Limits}.}
\label{fig:Hmutau-32}
\end{figure}

For LFV Higgs decays, our findings are illustrated in
~\reffis{fig:Hemu-12}-\ref{fig:Hmutau-32}. \reffi{fig:Hemu-12} displays
the $\br(h \to e \mu)$ as a function of $T_{12}^{E}$ (left column)
and $T_{12}^{\prime E}$ (right column) in the scenarios $M_{h}^{125}$
(upper row), $M_{h}^{125}(\tilde{\tau})$ (middle row), and
$M_{h}^{125}(\tilde{\chi})$ (lower row). The plots show that $\br(h
\to e \mu)$ can reach values of the $\order{10^{-17}}$
($\order{10^{-10}}$) for $T_{12}^{E}$ ($T_{12}^{\prime E}$) in
$M_{h}^{125}$ scenario while in $M_{h}^{125}(\tilde{\tau})$ they are
one-order-of-magnitude smaller compared to $M_{h}^{125}$ scenario. In
the $M_{h}^{125}(\tilde{\chi})$ scenario, the $\br(h \to e \mu)$
remain around $\order{10^{-17}}$ ($\order{10^{-18}}$) for $T_{12}^{E}$
($T_{12}^{\prime E}$). Interestingly, despite being more constrained
compared to $T_{12}^{E}$, the coupling $T_{12}^{\prime E}$ can result in
a branching ratio for $h \to e \mu$ that is seven orders of magnitude
larger compared to the $T_{12}^{ E}$ contributions in $M_{h}^{125}$ and
$M_{h}^{125}(\tilde{\tau})$ scenarios while still adhering to the
constraints on cLFV decays.
However, the obtained values are far below the current or
anticipated future limits (see below).
As anticipated, the $T_{12}^{\prime E}$
contributions increase with increasing values of $\tb$ and the most
substantial contributions are observed for the point P4, where $\tb=45$,
which can be attributed to the multiplication of NH $T_{ij}^{^\prime E}$
couplings by $\tb$. The results for $\br(h \to e \mu)$ as a
function of $T_{21}^{E}$ ($T_{21}^{\prime E}$) exhibit similar trends to
those for $T_{12}^{E}$ ($T_{12}^{\prime E}$), and are not presented
here.

\reffi{fig:Hetau-13} illustrates the $\br(h \to e \tau)$ as a
function of $T_{13}^{E}$ and $T_{13}^{^\prime E}$. The arrangement of
plots is same is in the previous figure. In the context of
$M_{h}^{125}$, the $\br(h \to e \tau)$ can be
three-order-of-magnitude larger for $T_{13}^{\prime E}$, reaching up to
$\order{10^{-2}}$, compared to $T_{13}^{E}$, where it can only reach up
to $\order{10^{-5}}$.
In the upper right plot we have indicated the bounds on
$T_{31}^{\prime E}$ that are placed by the current limits on
$\br(h \to e \tau)$, see \refta{tab:cLFV-BR-Limits}.
In the context of the $M_{h}^{125}(\tilde{\tau})$
scenario, for points P3, the contributions of $T_{13}^{E}$ to ${\rm
  BR}(h \to e \tau)$ are found at levels around
$\order{10^{-9}}$. Conversely, for P1 and P2, these contributions are
even smaller. In contrast, contributions from $T_{13}^{\prime E}$ reach
around $\order{10^{-4}}$ for P3, while maintaining values of
approximately $\order{10^{-10}}$ and $\order{10^{-9}}$ for P1 and P2,
respectively. For the $M_{h}^{125}(\tilde{\chi})$ scenario,
$T_{13}^{^\prime E}$ contributions can be two orders of magnitude larger
than $T_{13}^{ E}$ contributions for P3, while only a
one-order-of-magnitude difference is observed for other
points. \reffi{fig:Hetau-31} depicts the dependence of $\br(h \to e
\tau)$ on $T_{31}^{E}$ and $T_{31}^{^\prime E}$. The arrangement of the
plots follows the same pattern as in previous figures. Once again, in
the $M_{h}^{125}$ and $M_{h}^{125}(\tilde{\tau})$ scenarios, the
$T_{31}^{^\prime E}$ coupling can yield significantly large
contributions compared to $T_{31}^{E}$, even though the allowed range
for $T_{31}^{^\prime E}$ is considerably smaller than that of
$T_{31}^{E}$. For the points P3 and P4, the $\br(h \to e \tau)$ can
attain values up to $\order{10^{-2}}$ in $M_{h}^{125}$ scenario.
As in \reffi{fig:Hetau-13} we have indicated in the upper right
plot the limits placed on $T_{31}^{^\prime E}$ from the existing
limits on $\br(h \to e \tau)$, see \refta{tab:cLFV-BR-Limits}.
In the $M_{h}^{125}(\tilde{\chi})$ scenario, the results exhibit nearly
identical patterns, with $T_{31}^{^\prime E}$ resulting in a branching
ratio approximately one order of magnitude larger than that predicted by
$T_{31}^{E}$. 

The $\br(h \to \mu \tau)$ as a function of $T_{23}^{E}$ and
$T_{23}^{\prime E}$ is shown in \reffi{fig:Hmutau-23}. Again, the
arrangement of the plots is same as in the previous figures. The NH
trilinear coupling's effects are more pronounced in $M_{h}^{125}$ and
$M_{h}^{125}(\tilde{\tau})$ and less significant for
$M_{h}^{125}(\tilde{\chi})$ as was observed in previous figures.
As before, in the $M_h^{125}$ scenario the parameter space of
$T_{23}^{^\prime E}$ is restricted now from $\br(h \to \mu\tau)$ as
indicated in the upper right plot.
The relative difference between contributions from $T_{23}^{E}$ and
$T_{23}^{\prime E}$ mirrors that of $T_{13}^{E}$ and $T_{13}^{\prime
  E}$. Finally, we present the dependence of $\br(h \to \mu \tau)$ on
$T_{32}^{E}$ and $T_{32}^{\prime E}$ in \reffi{fig:Hmutau-32}. These
plots exhibit a similar pattern to the one discussed earlier in relation
to \reffi{fig:Hetau-31}.
Again, in the $M_h^{125}$ scenario bounds on $T_{32}^{^\prime E}$
are placed by the current bounds on $\br(h \to \mu\tau)$.
All the bounds on NH parameters found in this scenario from the existing
limits on $\br(h \to e \tau)$ and $\br(h \to \mu\tau)$ are summarized in
\refta{tab:Limits-LFVHiggs}. The limits from cLFV are given for
comparison and are identical to \protect\refta{tab:cLFV-Limits}.

Finally, for completeness, in \refta{tab:-Future-Limits}, we provide
future experimental bounds on 
cLFV and LFVHDs. These bounds are smaller by $\order{10^{-1}}$ compared
to the current experimental bounds.
Concerning the future anticipated bounds cLFV decays, in case of no
discovery, they will restrict severely the allowed parameter space of
the NH SSB terms. Concerning the future sensitivities on LFVHDs, it
becomes clear that the excesses observed by ATLAS, see also the next
subsection, can conclusively be tested. In particular in the $M_h^{125}$
scenario these bounds can restrict severely the allowed parameter space
of some of the NH SSB parameters.

In summary, NH trilinear couplings can yield significantly larger
contributions compared to holomorphic contributions, particularly in
$M_{h}^{125}$ and $M_{h}^{125}(\tilde{\tau})$. For
$M_{h}^{125}(\tilde{\chi})$, the predictions are small because $\mu$ is
smaller ($\mu = 180 \gev$) compared to $M_{h}^{125}$ or
$M_{h}^{125}(\tilde{\tau})$, where $\mu = 1000 \gev$. The
parameter $\mu$ appears in the mass matrix and in the couplings
multiplied by $Y_{ij}^{E}$. The non-zero values of $T_{ij}^{E}$ or
$T_{ij}^{\prime E}$ correspond to a non-zero $Y_{ij}^{E}$, which appears
as a prefactor of $\mu$ in the mass matrices and couplings.
Consequently, a larger $\mu$ will lead to larger LFV contributions.

\begin{table}[htb!]
\centerline{\begin{tabular}{|c|c|c|c|c|c|c|}
\hline\hline
  &  \multicolumn{2}{|c|}{P2} & \multicolumn{2}{|c|}{P3} &  \multicolumn{2}{|c|}{P4} \\ \hline
  &  cLFV &  LFVHD &  cLFV &  LFVHD &  cLFV &  LFVHD \\ \hline
$|T_{13}^{^\prime E}|$   &  1486 & 963 &  1468 & 477 & 1468 & 315 \\ 
$|T_{31}^{^\prime E}|$   &  1432 & 945 & 1414 & 477 & 1396 & 315\\ 
$|T_{23}^{^\prime E}|$   &  1684 & 801 & 1684 & 405 & 1666 & 261\\ 
$|T_{32}^{^\prime E}|$   &  1612 & 783 & 1612 & 387 & 1612 & 261\\ 
\hline\hline
\end{tabular}}
\caption{Constraints on $T_{ij}^{^\prime E}$ arising from LFV Higgs
  decays within the $M_{h}^{125}$ scenario. All the dimensionful
  quantities are in $\gev$. The limits from cLFV are given for
    comparison and are identical to \protect\refta{tab:cLFV-Limits}}.
\label{tab:Limits-LFVHiggs}
\end{table}

\clearpage

\begin{table}[htb!]
\centerline{\begin{tabular}{|c|c||c|c|}
\hline\hline
 cLFV Decays &  Limit &  LFVHD &  Limit   \\ 
 \hline\hline
$\br(\mu \to e \gamma)$ & $6.0 \times 10^{-14}$ \cite{MEGII:2018kmf}&  $\br(h \to e \mu)$ & $1.2 \times 10^{-5}$\cite{Qin:2017aju}    \\ 
$\br(\tau \to e \gamma)$ &   $9.0\times 10^{-9}$\cite{Belle-II:2018jsg} &$\br(h \to e \tau)$ & $1.6 \times 10^{-4}$\cite{Qin:2017aju}\\ 
 $\br(\tau \to \mu \gamma)$ & $6.9 \times 10^{-9}$\cite{Belle-II:2018jsg} & $\br(h \to \mu \tau)$ & $1.4 \times 10^{-4}$\cite{Qin:2017aju}  \\  
\hline\hline
\end{tabular}}
\caption{Future upper bounds on the cLFV decays and LFV Higgs
  decays~\cite{MEGII:2018kmf,Belle-II:2018jsg,Qin:2017aju}.}
\label{tab:-Future-Limits}
\end{table}


\subsection{The ATLAS excess}
\label{sec:atlas}

Recently ATLAS reported an excess~\cite{ATLAS:2023mvd}
in their searches for $h \to e \tau$ and $h \to \mu\tau$. Together these
two channels show an excess larger than $2\,\si$. Their best-fit values
of $\br(h \to e \tau) \approx \br(h \to \mu\tau) \approx 0.1\%$ is not
excluded by the corresponding CMS limits~\cite{CMS:2021rsq},
$\br(h \to e\tau) < 0.22\%$ and $\br(h \to \mu\tau) < 0.15\%$ at the
95\% C.L.~limit.
In the previous subsection we have demonstrated that the NHSSM can yield
values of these two BRs that are in the ballpark of the ATLAS excess for
each of the two BRs individually. In this section 
we demonstrate that the NHSSM can accommodate both
excesses simultaneously without being in conflict with other
experimental limits. 

\begin{figure}[htb!]
\centering
\includegraphics[width=7cm,keepaspectratio]{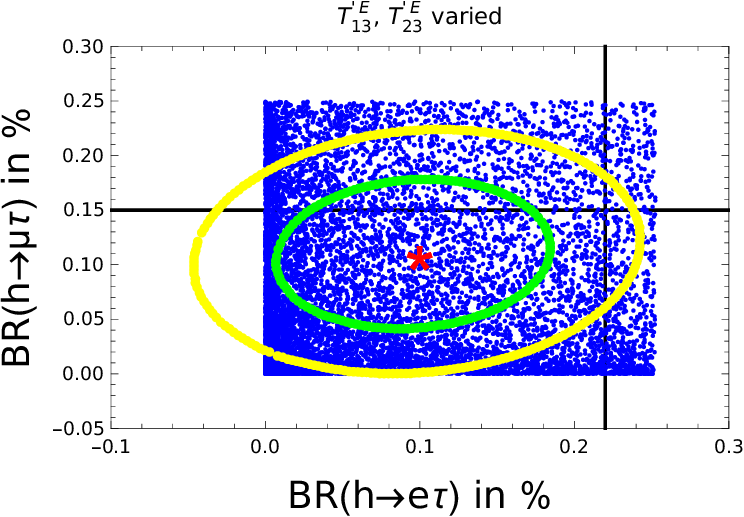}
\hspace{0.02\textwidth}
\includegraphics[width=7cm,keepaspectratio]{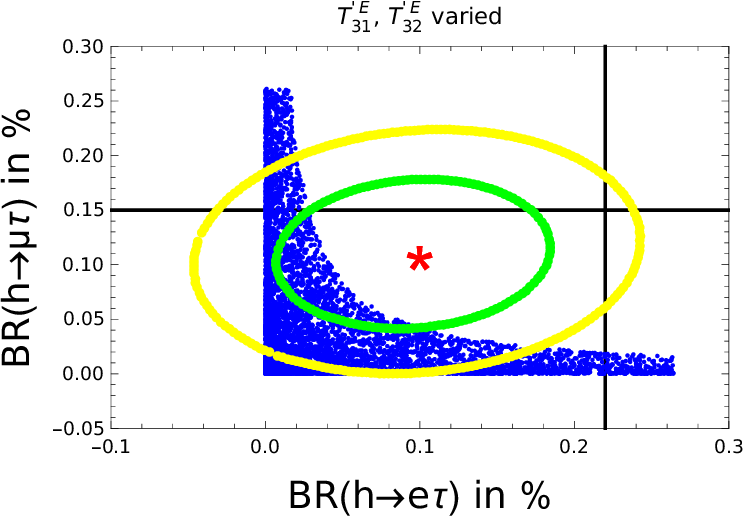}\\[0.5em]
\vspace{0.03\textwidth}
\includegraphics[width=7.0cm, height=7.0cm]{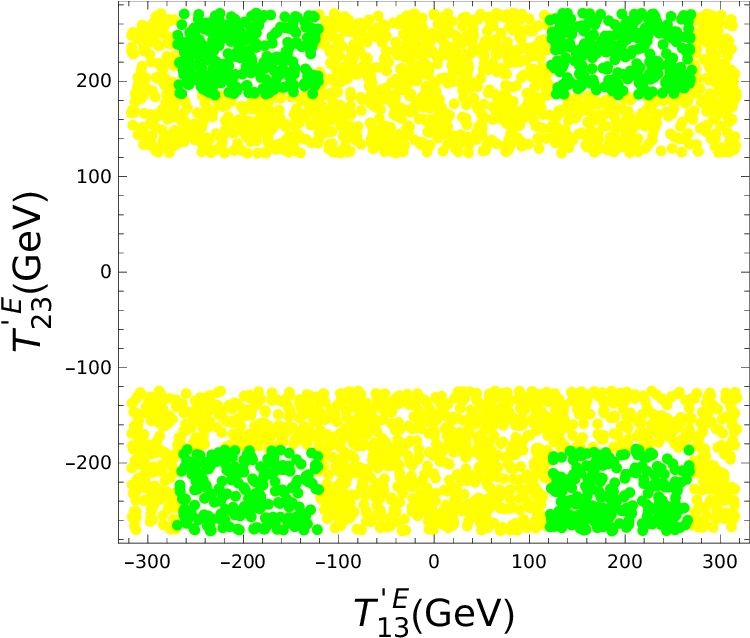}
\hspace{0.02\textwidth}
\includegraphics[width=7.0cm, height=7.0cm]{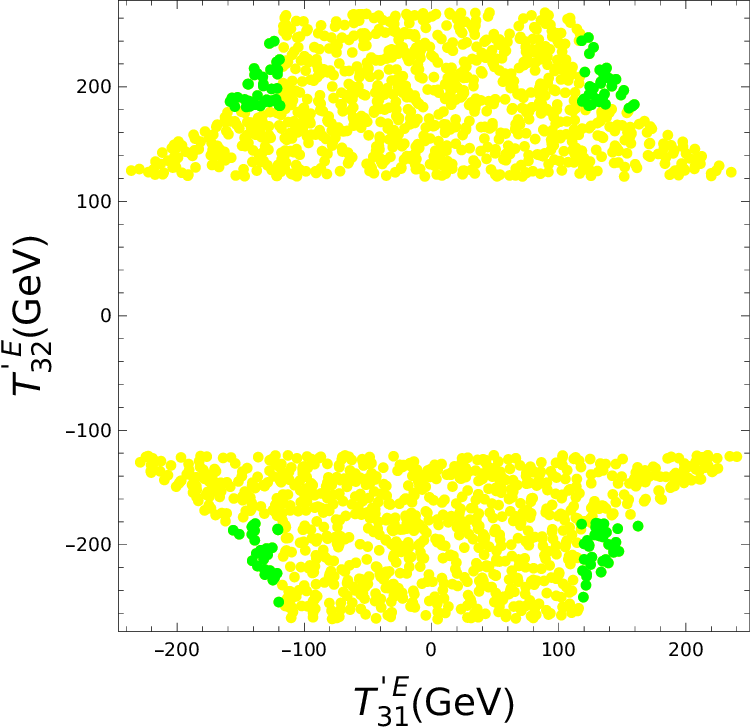}\\
\caption{
  Upper row:
  the $\br(h \to \mu\tau)$--$\br(h \to e\tau)$ plane showing the ATLAS
  excess (ellipses), the CMS bounds (horizontal and vertical solid
  lines), and our predictions in the $M_h^{125}$ scenario for P4
  ($\MA = 2500 \gev$, $\tb = 45$) with both
  $T_{13}^{^\prime E}$ and $T_{23}^{^\prime E}$ varied independently
  (upper-left plot)
  and $T_{31}^{^\prime E}$ and $T_{32}^{^\prime E}$ (upper-right plot).
  Lower row:
  the points found in the $1\,\si$ and $2\,\si$ ellipses of the ATLAS
  excesses are shown in the $T_{13}^{^\prime E}$--$T_{23}^{^\prime E}$ (lower-left) and $T_{31}^{^\prime E}$--$T_{32}^{^\prime E}$ (lower-right)
  plane as yellow and green dots, respectively.}
\label{fig:Contour-br-atlas}
\end{figure}

In the upper plots of \reffi{fig:Contour-br-atlas} we show the
$\br(h \to \mu\tau)$--$\br(h \to e\tau)$ plane indicating the ATLAS 
excess (ellipses) and the CMS bounds (horizontal and vertical solid
lines).
The red stars denote the best-fit point from ATLAS~\cite{ATLAS:2023mvd}.
Within the $M_h^{125}$ scenario for P4 (i.e.\ $\MA = 2500 \gev$,
$\tb = 45$), we randomly scanned $T_{13}^{^\prime E}$ ($T_{31}^{^\prime E}$)
and $T_{23}^{^\prime E}$ ($T_{32}^{^\prime E}$), yielding the upper left (right) plot.
For all points we took into account the current bounds from cLFV.
One can observe that the $1\,\si$ ellipse of the ATLAS excesses is well
populated for the scan varying $T_{13}^{^\prime E}$ and $T_{23}^{^\prime E}$
(upper left plot), 
while only the ``lower values'' of the ellipse are populated in the scan of
$T_{31}^{^\prime E}$ and $T_{32}^{^\prime E}$ (upper right plot).

In the lower row of \reffi{fig:Contour-br-atlas} we show the points
found in the $1\,\si$ and $2\,\si$ ellipses in the upper row as green
and yellow points, respectively, in the
$T_{13}^{^\prime E}$--$T_{23}^{^\prime E}$ plane (left plot) and in the
$T_{31}^{^\prime E}$--$T_{32}^{^\prime E}$ plane (right plot). 
All points are below the respective CMS
limits indicated as solid horizontal and vertical lines in the upper
plots. One can observe that the NH contribution to $\br(h \to \mu\tau)$
and $\br(h \to e \tau)$ are invariant under the signs of $T_{ij}^{\prime E}$.
The points inside the $1\,\si$ ellipses are found for
$100 \gev \lsim |T_{13,31}^{\prime E}| \lsim 250 \gev$ and
$180 \gev \lsim |T_{23,32}^{\prime E}| \lsim 280 \gev$. 

Overall, we conclude that the NHSSM can 
describe the observed excesses well, while being in agreement with the
existing experimental bounds. 

\section{Conclusions}
\label{sec:conclusions}

The MSSM including nonholomorphic soft SUSY-breaking terms
(NHSSM) is a well motivated model going beyond the SM and extending
non-trivially the MSSM. Within the NHSSM, we investigated the effects of
lepton flavor violating terms.  
In a first step we applied the current bounds from charged lepton flavor
violating (cLFV) decays, $l_i \to l_j \ga$ ($i,j=1,2,3$, $i > j$,
$l_{1,2,3} = e, \mu, \tau$) to set
limits on the NH flavor violating soft SUSY-breaking parameters,
$T_{ij}^{^\prime E}$ ($i,j = 1,2,3$).
Our approach, systematically altering one parameter at
a time, provides a clear indication of the NH contributions to the cLFV
decays.

In a second step we calculated LFV Higgs decays (LFVHD),
$h \to l_i l_j$ ($i,j = 1,2,3$, $i \neq j$), taking into account the
bounds on $T_{ij}^{^\prime E}$ obtained from cLFV decays. We found that
the $\br(h \to l_i l_j)$ can reach values far above the corresponding SM
values (as was known from the literature), but even reaching and
exceeding the current bounds from CMS and ATLAS. 

Interestingly, recently ATLAS reported an excess~\cite{ATLAS:2023mvd}
in their searches for $h \to e \tau$ and $h \to \mu\tau$. Together these
two channels show an excess larger than $2\,\si$. Their best-fit values
of $\br(h \to e \tau) \approx \br(h \to \mu\tau) \approx 0.1\%$ is not
excluded by the corresponding CMS limits~\cite{CMS:2021rsq},
$\br(h \to e\tau) < 0.22\%$ and $\br(h \to \mu\tau) < 0.15\%$ at the
95\% C.L.~limit. We varied independently the parameters
$T_{13}^{^\prime E}$, $T_{23}^{^\prime E}$ or $T_{31}^{^\prime E}$, $T_{32}^{^\prime E}$
to test whether the ATLAS excess can be described within the NHSSM.
We demonstrate that the model can accommodate both
excesses without being in conflict with other experimental limits.
If these decays are eventually observed experimentally, they could
potentially serve as a distinctive signature of the NH scenarios
and determine some of the NH parameters.
Conversely, the limits on the LFVHDs can
restrict the allowed parameter space for the NH SSB terms in the NHSSM. 


\subsection*{Acknowledgments}

The work of M.R.\ was supported by HEC Pakistan, under NRPU grant 20-15867/NRPU/ R\&D/HEC/2021.
The work of S.H.\ has received financial support from the
PID2022-142545NB-C21 funded by MCIN/AEI/10.13039/501100011033/ FEDER, UE
and in part by the grant IFT Centro de Excelencia Severo Ochoa CEX2020-001007-S
funded by MCIN/AEI/ 10.13039/501100011033.


\end{document}